\documentclass[prb,twocolumn,superscriptaddress]{revtex4-2}
\usepackage{amsmath}
\usepackage{color}
\usepackage{bbm}
\usepackage{amssymb}
\usepackage{epsfig}
\usepackage{multirow}
\usepackage{amsbsy}
\usepackage{array}
\usepackage{diagbox}
\usepackage{bm}
\usepackage{tikz}
\usepackage{extarrows}
\usepackage{graphicx}
\usepackage{appendix}
\usepackage{txfonts}
\usepackage[normalem]{ulem}
\usepackage[colorlinks=true,linkcolor=blue,citecolor=blue,urlcolor=blue,bookmarks=false]{hyperref}

\definecolor{orange}{rgb}{1,0.5,0}
\definecolor{usubeni}{RGB}{232,122,144}

\newcommand{\ket}[1]{\left|{#1}\right\rangle}

\newcommand{\ii}[0]{\mathrm{i}}
\newcommand{\ee}[0]{\mathrm{e}}

\begin{document}

\title{Symmetry-protected topological order identified via Gutzwiller-guided density-matrix-renormalization-group: $\mathrm{SO}(n)$ spin chains}

\author{Pei-Yuan Cai}
\affiliation{Institute of Physics, Chinese Academy of Sciences, Beijing 100190, China}
\affiliation{University of Chinese Academy of Sciences, Beijing 100190, China}

\author{Hui-Ke Jin}
\email{jinhk@shanghaitech.edu.cn}
\affiliation{State Key Laboratory of Quantum Functional Materials, School of Physical Science and Technology, ShanghaiTech University, Shanghai 201210, China}

\author{Yi Zhou}
\email{yizhou@iphy.ac.cn}
\affiliation{Institute of Physics, Chinese Academy of Sciences, Beijing 100190, China}

\date{\today}

\begin{abstract}
We present a comprehensive study of topological phases in the SO($n$) spin chains using a combination of analytical parton construction and numerical techniques. For even $n=2l$, we identify a novel SPT$^2$ phase characterized by two distinct topological sectors, exhibiting exact degeneracy at the matrix product state (MPS) exactly solvable point. Through Gutzwiller-projected mean-field theory and density matrix renormalization group (DMRG) calculations, we demonstrate that these sectors remain topologically degenerate in close chains throughout the SPT$^2$ phase, with energy gaps decaying exponentially with system size. For odd $n=2l+1$, we show that the ground state remains unique in close chains. We precisely characterize critical states using entanglement entropy scaling, confirming the central charges predicted by conformal field theories. Our results reveal fundamental differences between even and odd $n$ cases, provide numerical verification of topological protection, and establish reliable methods for studying high-symmetry quantum systems. The Gutzwiller-guided DMRG is demonstrated to be notably efficient in targeting specific topological sectors.
\end{abstract}

\maketitle

\section{Introduction}\label{sec:introduction}

One-dimensional quantum models have long served as a fertile ground for exploring fundamental concepts in condensed matter physics. These systems exhibit a wide range of fascinating phenomena, such as exactly solvable points and symmetry-protected topological orders. The interplay between symmetry and topology in these systems has led to the discovery of unique quantum states that cannot be characterized by local order parameters, but instead by their global topological properties.

Paradigmatic one-dimensional models such as the Heisenberg chain~\cite{bethe1931theorie,haldane1983continuum,haldane1983nonlinear,Majumdar1969}, the transverse-field Ising (TFI) chain~\cite{pfeuty1970one,sachdev1999quantum,vidal2003entanglement}, the Affleck-Kennedy-Lieb-Tasaki (AKLT) chain~\cite{affleck1988valence,affleck2004rigorous}, and the Kitaev chain~\cite{kitaev2001unpaired} have demonstrated how this interplay produces states protected by global symmetries. These findings have laid the foundation for modern studies of symmetry-protected topological (SPT) phases~\cite{gu2009tensor,Chen2012,Chen2013,Pollmann2010}. Unlike conventional phases, SPT phases are distinguished by their protected edge states that remain stable under perturbations preserving the system's symmetries~\cite{gu2009tensor}.

The bilinear-biquadratic (BBQ) spin chain, initially introduced as an intuitive advancement of the Heisenberg model to include higher-order spin correlations~\cite{bbq1,bbq2}, has evolved into a fundamental framework for examining numerous quantum phases, including traditional ferro- and antiferromagnetism, as well as spin-nematic and SPT orders. For one-dimensional spin-1 systems, early studies disclosed the renowned Haldane gap along with symmetry-protected edge modes~\cite{bbq3,bbq1d1,bbq1d2}. This, in turn, offered a microscopic setting for understanding quantum spin liquids and SPT order in one dimension. A significant generalization is obtained by imposing an SO($n$) symmetry, which promotes the spin-1 chain to an $\mathrm{SO}(n)$ BBQ chain with independently tunable bilinear and biquadratic couplings~\cite{liu2010fermionic,tu2011effective,tu2008class,ragone2024so}. At special parameter values this model is exactly solvable or admits controlled field-theory descriptions, enabling the identification of critical points and universal properties~\cite{reshetikhin1983method,affleck1988valence,Pollmann2010,orus2011phase,uimin1970one}.  Systematic analytical and numerical studies for both even and odd $n$~\cite{Scalapino1998,kolezhuk1998non,itoi2000phase,azaria2000effect,tu2008string,alet2011quantum,affleck19912n,ragone2024so} have uncovered pronounced variations in ground-state degeneracy and topological properties, underscoring the model’s sensitivity to symmetry and spin magnitude. Nevertheless, a unified understanding of how these topological characteristics evolve across the full parameter space remains elusive.  Given that SO($n$) BBQ chains can now be emulated in quantum magnets and ultracold-atom platforms, resolving this open question is both fundamentally and experimentally timely.

To address this unsolved issue, we employ the Gutzwiller projected wave function approach, which has proven valuable for studying strongly correlated systems~\cite{gros1989physics, lee2006doping, zhou2017quantum}. This method offers a powerful framework for constructing ground states by imposing local constraints. Inspired by previous applications to spin- $S\ge{}1$ chains~\cite{liu2010fermionic,liu2012gutzwiller,liu2014gutzwiller}, we extend this methodology to study the $\mathrm{SO}(n)$ BBQ chain. To enhance our analytical and numerical investigations, we incorporate the matrix product operator-matrix product state (MPO-MPS) framework~\cite{wu2020tensor, jin2020efficient}, which enables efficient construction of BCS mean-field states in matrix product state (MPS) form~\cite{Jin2022MPS}. The Gutzwiller projection can then be realized by applying local projectors to each local tensor of the MPS. This approach allows us to construct trial wave functions and optimize them numerically using the density matrix renormalization group (DMRG) method~\cite{white1992density, schollwock2011density, jin2021density}.

In this paper, we focus on the $\mathrm{SO}(n)$ spin BBQ chain using parton construction and fermionic mean-field theory. Our key finding is the discovery of a unique SPT$^2$ phase for even values of $n$, which hosts \emph{two-fold topologically degenerate ground states} in a close chain. In particular, at the so-called ``MPS exactly solvable point"~\cite{tu2008class}, which is the $\mathrm{SO}(n)$ generalization of the AKLT point, this topological degeneracy becomes exact due to the enlarged symmetry. In parton language, these two topologically degenerate ground states arise from the Gutzwiller-projected mean-field ground states under anti-periodic boundary conditions (APBC) and periodic boundary conditions (PBC). In contrast, when $n$ is odd, only one of the mean-field ground states survives the Gutzwiller projection, resulting in a unique ground state at the $\mathrm{SO}(n=2l+1)$ MPS exactly sovable point. Additionally, we show that the degeneracy of ground states under open boundary conditions (OBC) can be precisely counted using the Gutzwiller-projected wave function approach. 

To further understand these ground states, we combine analytical techniques with numerical simulations using the Gutzwiller-projected DMRG approach. Away from the MPS exactly solvable point, this approach allows us to identify the two distinct topological sectors and confirm that the energy gap between them decays exponentially with increasing system size $L$, verifying their topological degeneracy in the thermodynamic limit. Furthermore, utilizing the MPS representation of wave functions, we compute the central charges at several critical points using the Calabrese-Cardy formula~\cite{calabrese2009entanglement}.

The rest of this paper is organized as follows. In Section~\ref{sec:model}, we introduce the $\mathrm{SO}(n)$ spin BBQ model and outline its parton construction as $n$ decoupled Kitaev chain, providing the theoretical foundation for our analysis. Section~\ref{sec:MPSpoint}, with the parton construction method, we elucidate the even-odd effect of $n$ at the MPS exactly solvable point, identify a $\mathrm{Z}_2\cong\mathrm{O}(n)/\mathrm{SO}(n)$ symmetry breaking for even $n$, and complete the degeneracy counting under OBC. In section~\ref{sec:results}, using Gutzwiller-guided DMRG, we report our numerical results beyond the MPS exactly solvable point, including topological ground-state degeneracy in the SPT$^2$ phases and the central charges for several critical states. Section~\ref{sec:sum} provides a summary and discussion of our findings and their implications for future researchs.

\section{Model and Formulation}\label{sec:model}

The Hamiltonian of the SO($n$) spin BBQ model is defined on a chain of length $L$:
\begin{equation}\label{eq:HBBQ}
H=\sum_{i=1}^L\left[J \sum_{a<b} \mathbf{L}_i^{a b} \mathbf{L}_{i+1}^{a b}+K\left(\sum_{a<b} \mathbf{L}_i^{a b} \mathbf{L}_{i+1}^{a b}\right)^2\right],
\end{equation}
where $\mathbf{L}_i^{ab}$ ($1\leq a<b\leq n$) are the $n(n-1)/2$ generators of the $\mathrm{SO}(n)$ Lie algebra, satisfying the commutation relation:
$$\left[\mathbf{L}_i^{a b}, \mathbf{L}_j^{c d}\right]=\ii\delta_{ij}\left(\delta^{a c} \mathbf{L}_i^{b d}-\delta^{a d} \mathbf{L}_i^{b c}-\delta^{b c} \mathbf{L}_i^{a d}+\delta^{b d} \mathbf{L}_i^{a c}\right).$$ 
Higher order terms in Eq.~\eqref{eq:HBBQ} vanish as they can be expressed as combinations of the bilinear and biquadratic terms~\cite{tu2008class}. The local states $\lvert{m_i^a}\rangle$ ($1\leq a\leq n$) transform according to the $\mathrm{SO}(n)$ rotation rule:
\begin{equation}\label{eq:son_rotation}
\mathbf{L}_i^{a b}\lvert{m_i^c}\rangle=\ii \delta_{b c}\lvert{m_i^a}\rangle-\ii \delta_{a c}\lvert{m_i^b}\rangle.
\end{equation}
Throughout our discussion, we assume periodic boundary conditions $\mathbf{L}^{ab}_{L+1} = \mathbf{L}^{ab}_1$ unless explicitly stated otherwise. The real coefficients $J$ and $K$ are conventionally parameterized as \begin{equation}\label{eq:JKtheta}
J=\cos\theta\quad\mbox{and}\quad K=\sin\theta
\end{equation}
to obtain a normalized single-parameter model. We will use both notations interchangeably in this paper.

\subsection{Parton Representation}

A key insight of this paper is that parton construction provides a straightforward framework for understanding several important phases in the BBQ model. We develop a fermionic parton theory by introducing $n$ species of fermionic operators at each lattice site $i$:
$$\mathbf{a}^\dagger_{i}=\left(a^\dagger_{i1},a^\dagger_{i2},\cdots,a^\dagger_{in}\right),$$ 
where the index $\alpha=(1,2,\cdots,n)$ denotes the flavor of $a^\dagger_{i\alpha}$.
The $\mathrm{SO}(n)$ generators then can be represented as:
\begin{equation}
\mathbf{L}_{i}^{ab}=\mathbf{a}^\dagger_{i}\mathbf{L}^{ab}\mathbf{a}_{i}=\sum_{\alpha,\beta} a_{i\alpha}^\dagger \big(\mathbf{L}^{ab}\big)_{\alpha\beta} a_{i\beta}, 
\end{equation}
where $(\mathbf{L}^{ab})_{\alpha\beta}$ is the $n\times{n}$ matrix representation of $\mathbf{L}^{ab}$ in the $\{|m^a\rangle\}$ basis. This parton representation enlarges the $n$-dimensional local Hilbert space to a $2^n$-dimensional one. To recover the physical Hilbert space, we must impose the local constraint:
$$\sum_{\alpha}a^\dagger_{i\alpha}a_{i\alpha}=1$$ at each lattice site $i$.

We define the $\mathrm{SO}(n)$-singlet bond operators:
\begin{equation}\label{chi_and_del_operators}
\hat{\chi}_{ij} = \sum_{\alpha=1}^n a_{i\alpha}^\dagger a_{j\alpha}, \quad \hat{\Delta}_{ij}=\sum_{\alpha=1}^n a_{i\alpha}a_{j\alpha}, 
\end{equation}
which allow us to rewrite the BBQ Hamiltonian in Eq.~\eqref{eq:HBBQ} as:
\begin{equation}\label{HBBQ_in_chidel}
H = \sum_{\left\langle i,j \right\rangle}\left\{ -J\hat{\chi}_{ij}^\dagger \hat{\chi}_{ij} + \left[(n-2)K-J\right]\hat{\Delta}_{ij}^\dagger \hat{\Delta}_{ij}+(K+J)\right\}.
\end{equation}
The proof of Eq.~(\ref{HBBQ_in_chidel}) is provided in Appendix A. This formulation immediately reveals an important critical point at $K/J=\tan\theta=1/(n-2)$, where the $\hat{\Delta}_{ij}^\dagger \hat{\Delta}_{ij}$ terms vanish and the model exhibits an enhanced $\mathrm{SU}(n)$ symmetry. This critical point corresponds precisely to the Uimin-Lai-Sutherland (ULS) model, which is integrable and solvable using the Bethe-ansatz method~\cite{uimin1970one, lai1974lattice, sutherland1975model}.

\subsection{Mean-Field Theory and Gutzwiller Projection}

\begin{figure}[tb]
\centering
\includegraphics[width=1.0\linewidth]{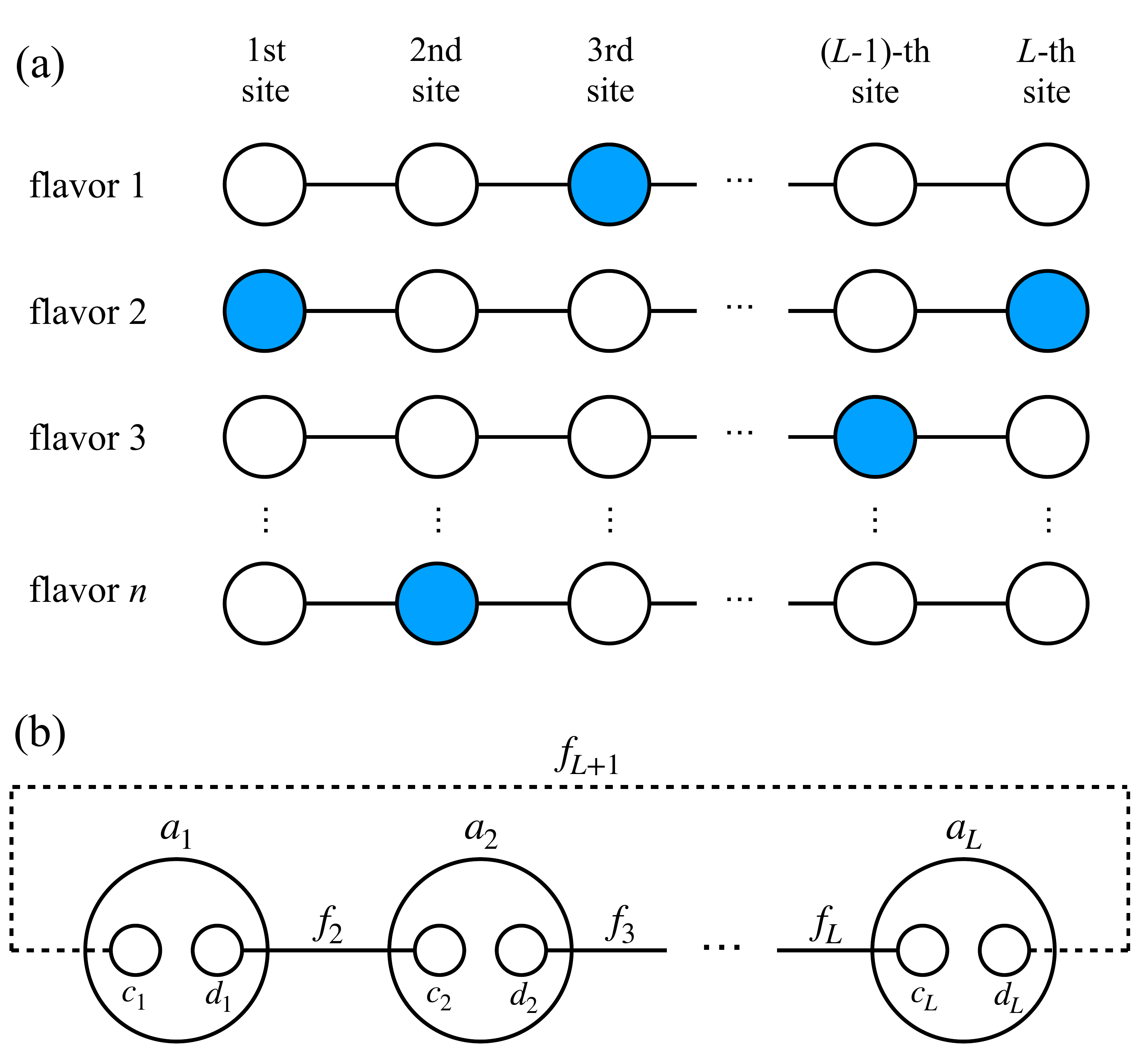}
\caption{(a) The mean-field Hamiltonian can be expressed as $n$ decoupled copies of Kitaev chains. The blue-filled sites of the Kitaev chains correspond to occupied parton states, while the hollow sites indicate unoccupied parton states. The figure illustrates a possible configuration of $n$ Kitaev chains under the $1/n$ filling constraint. (b) A single Kitaev chain of length $L$. The Majorana fermions $c_i$ and $d_i$ are recombined into bond operators $f_j$. The dashed line represents the boundary term of the bond operator. }
\label{fig:fig1}
\end{figure}

By introducing mean-field parameters $\chi=\langle\hat{\chi}_{ij}\rangle$, $\Delta=\langle\hat{\Delta}_{ij}\rangle$, and a Lagrange multiplier $\lambda$ as the chemical potential, the BBQ Hamiltonian in Eq.~\eqref{HBBQ_in_chidel} can be decoupled into $n$ copies of Kitaev chains[see Fig.~\ref{fig:fig1}~(a)]:
\begin{equation}\label{eq:HMF}
H_{\mathrm{MF}} =\sum_{\alpha=1}^{n} H^\alpha_{\mathrm{K}}.
\end{equation}
Here, $H^\alpha_{\mathrm{K}}$ is the Hamiltonian of a single Kitaev chain with parton flavor $\alpha$:
\begin{equation}
H^\alpha_{\mathrm{K}} = \sum_{i} \left[\left(-\tilde{\chi} a_{i\alpha}^{\dagger} a_{i+1, \alpha}+\tilde{\Delta} a_{i\alpha} a_{i+1, \alpha}+\mathrm{h.c.}\right) +\lambda a_{i\alpha}^{\dagger} a_{i\alpha}\right],
\end{equation}
where we introduce the re-scaled mean-field parameters as $\tilde{\chi}=J\chi$ and $\tilde{\Delta}=[J-(n-2)K]\Delta$.

For a given set of parameters $\{\tilde{\chi}, \tilde{\Delta}, \lambda\}$, the mean-field ground state for flavor $\alpha$, denoted as $|\Psi_{\mathrm{MF}}^\alpha\rangle$, can be obtained by diagonalizing $H^\alpha_{\mathrm{K}}$. The trial ground state wave function for the $\mathrm{SO}(n)$ spin BBQ model is constructed as:
\begin{equation}\label{eq:mf_to_trail}
|\Psi_{\mathrm{trial}}\rangle\equiv \mathbf{P}_{\mathrm{G}}|\Psi_{\mathrm{MF}}\rangle=\mathbf{P}_{\mathrm{G}}\prod_{\alpha=1}^{n}|\Psi_{\mathrm{K}}^{\alpha}\rangle,
\end{equation}
where $\mathbf{P}_\mathrm{G}$ is the Gutzwiller projection operator that imposes the single-occupancy constraint $\sum_{\alpha=1}^{n} a_{i\alpha}^\dagger a_{i\alpha} = 1$ at each lattice site $i$.

Due to the single-occupancy constraint, the fermion number parity of the Gutzwiller-projected state must be $(-1)^{L}$, where $L$ is the length of the $\mathrm{SO}(n)$ spin BBQ chain. Since the Gutzwiller projection does not alter the fermion number parity of a parity-conserving state, the parity of $|\Psi_{\text{trial}}\rangle$ and $\prod_{\alpha=1}^{n}|\Psi_{\mathrm{K}}^{\alpha}\rangle$ must be identical. Denoting $\mathrm{Z}_2^\alpha$ as the fermion number parity of the $\alpha$-flavor mean-field state, we arrive at an important constraint:
\begin{equation}\label{eq:even_parity_requirement}
\prod_{\alpha=1}^{n} \mathrm{Z}_2^{\alpha} = (-1)^{L}.
\end{equation}
Consequently, the total parity of $|\Psi_{\mathrm{MF}}\rangle$ must be $(-1)^L$; otherwise, it will be eliminated by the Gutzwiller projection. 

\subsection{Example models: $n=3$ and $n=4$}

The SO($n$) spin BBQ models can be realized in various physical systems. For instance, the $n=3$ model can be implemented in spin $S=1$ systems that exhibit both dipolar and quadrupolar spin interactions. Meanwhile, the $n=4$ model describes two-orbital systems involving spin-orbital interactions, where the symmetry can be enhanced to SU(4) or SO(6) through carefully tuned interactions.

\subsubsection{$n=3$: spin $S=1$ systems}
For $n=3$, the Hamiltonian can be represented by spin-1 operators: 
\begin{equation}\label{SO3BBQ}
H_{\mathrm{SO}(3)} = \sum_{i} \left[J \mathbf{S}_i\cdot\mathbf{S}_{i+1}+K\left( \mathbf{S}_i\cdot\mathbf{S}_{i+1}\right)^2\right],
\end{equation}
where $\mathbf{S}_i = (S_i^x, S_i^y, S_i^z)$ represents the spin operators at site $i$. 

This $H_{\mathrm{SO}(3)}$ is equivalent to Eq.~\eqref{eq:HBBQ} when we identify the generators as $\mathbf{L}^{12} = -S^z$, $\mathbf{L}^{13} = S^y$, and $\mathbf{L}^{23} = -S^x$. For spin-1 systems, it is more natural to use the standard basis $\{\ket{1}, \ket{0}, \ket{-1}\}$. At any given site $i$, we can define three single-occupied local parton states with different flavors as:
\[
|1\rangle=a_1^{\dagger}|\mathrm{vac}\rangle_a, \quad|0\rangle=a_0^{\dagger}|\mathrm{vac}\rangle_a, \quad|-1\rangle=a_{-1}^{\dagger}|\mathrm{vac}\rangle_a, 
\]
where $a_\alpha^\dagger$ is the fermionic creation operator for a parton with flavor $\alpha$, and $\ket{\mathrm{vac}}_a$ is the parton vacuum state. For the $n=3$ case, the spin-1 operators can be represented using three species of fermions (partons): 
$$
S_i^a=\sum_{\alpha,\beta=1,0,-1} a_{i \alpha}^{\dagger} I_{\alpha \beta}^a a_{i \beta}, 
$$
where $I^a_{\alpha\beta}=\left\langle \alpha |S^a|\beta\right\rangle$ is the matrix element of $S^a$ in the parton representation. A local constraint $\sum_\alpha a_{i \alpha}^{\dagger} a_{i \alpha}=1$ must be imposed to preserve the physical states.

To reveal the $\mathrm{SO}(3)$ symmetry of this model in alignment with the general Hamiltonian form, we transform to an $\mathrm{SO}(3)$-symmetric basis $\{\ket{x}, \ket{y}, \ket{z}\}$. The corresponding parton creation operators transform as:
$$
a_x^\dagger=\frac{1}{\sqrt{2}}(a_{-1}^\dagger-a_1^\dagger), \quad a_y^\dagger=\frac{\ii}{\sqrt{2}}(a_1^\dagger+a_{-1}^\dagger), \quad a_z^\dagger=a_0^\dagger.
$$
Using these operators, we can define the fermion hopping operator $\hat{\chi}_{ij}=\sum_{\alpha=x, y, z} a_{i \alpha}^{\dagger} a_{j \alpha}$ and the singlet pairing operator $\hat{\Delta}_{ij}=\sum_{\alpha=x, y, z} a_{i \alpha} a_{j \alpha}$ according to Eq.~(\ref{chi_and_del_operators}), allowing us to rewrite the Hamiltonian as:
\begin{equation}
H_{\mathrm{SO}(3)} = \sum_{\langle i, j\rangle} \left[ -J \hat{\chi}_{i j}^{\dagger} \hat{\chi}_{i j} + \left(K-J\right)\hat{\Delta}_{i j}^{\dagger} \hat{\Delta}_{i j}\right], 
\end{equation}
which is the $n=3$ case of Eq.~(\ref{HBBQ_in_chidel}).

\subsubsection{$n=4$: spin-orbital systems}
The Lie group $\mathrm{SO}(4)$ can be factorized as $\mathrm{SO}(4)\simeq \mathrm{SU}(2)\times \mathrm{SU}(2)$. Thus, we can consider a spin-orbital system with $S=T=1/2$ to implement the $\mathrm{SO}(4)$ vectors and generators. We introduce the basis states:
\begin{equation*}
\begin{aligned}
\left|n^{1}\right\rangle&=\frac{\ee^{ + \ii \pi / 4}}{\sqrt{2}}(\ket{\uparrow\uparrow} - \ket{\downarrow\downarrow}), \quad \left|n^{2}\right\rangle=\frac{\ee^{ - \ii \pi / 4}}{\sqrt{2}}(\ket{\uparrow\uparrow} + \ket{\downarrow\downarrow}), \\
\left|n^{3}\right\rangle&=\frac{\ee^{ - \ii \pi / 4}}{\sqrt{2}}(\ket{\uparrow\downarrow} - \ket{\downarrow\uparrow}), \quad \left|n^{4}\right\rangle=\frac{\ee^{ + \ii \pi / 4}}{\sqrt{2}}(\ket{\uparrow\downarrow} + \ket{\downarrow\uparrow}), \\
\end{aligned}
\end{equation*}
where $\ket{\sigma,\tau} \in \{\ket{\uparrow\uparrow}, \ket{\uparrow\downarrow}, \ket{\downarrow\uparrow}, \ket{\downarrow\downarrow}\}$ represents the natural basis denoting the spin and orbital directions. As suggested in Ref.~\cite{tu2008class}, the $\mathrm{SO}(4)$ generators can be defined as:
\begin{equation*}
\begin{aligned}
& \mathbf{L}^{12}=-T^z-S^z, \quad &\mathbf{L}^{13}=T^x-S^x, \quad &\mathbf{L}^{14}=-T^y-S^y, \\
& \mathbf{L}^{23}=T^y-S^y, \quad &\mathbf{L}^{24}=T^x+S^x, \quad &\mathbf{L}^{34}=T^z-S^z. 
\end{aligned}
\end{equation*}
With the $\mathrm{SO}(4)$-symmetric basis $\{\ket{n^\alpha}\}$ established, we can express the Hamiltonian as:
\begin{equation}\label{so4_hamiltonian_in_chidel}
H_{\mathrm{SO(4)}} = \sum_{\langle i, j\rangle} \left[ -J \hat{\chi}_{i j}^{\dagger} \hat{\chi}_{i j} + \left(2K-J\right)\hat{\Delta}_{i j}^{\dagger} \hat{\Delta}_{i j}\right], 
\end{equation}
where the hopping and pairing operators are $\hat{\chi}_{ij}=\sum_{\alpha=1}^4 a_{i \alpha}^{\dagger} a_{j \alpha}$ and $\hat{\Delta}_{ij}=\sum_{\alpha=1}^4 a_{i \alpha} a_{j \alpha}$. This is the $n=4$ case of Eq.~(\ref{HBBQ_in_chidel}).

\section{MPS exactly solvable points}~\label{sec:MPSpoint}

The SO$(n>3)$ spin chains possess exact ground states at the point of $\tan\theta=1/n$, known as MPS exactly solvable point.
The exact solvability is revealed by the fact that, at this point, the ground states of $\mathrm{SO}(n=2l+1)$ chains can be expressed as translationally invariant MPSs. Each local MPS tensor is formulated using $(2l+1)$ Gamma matrices~\cite{tu2008class} that generate the $2^{l}$-dimensional representation of the Clifford algebra. Meanwhile, the ground states of the $\mathrm{SO}(n=2l)$ chain are also constructed using the generators of this $2^{l}$-dimensional representation of the Clifford algebra. Since there are two distinct $\mathrm{SO}(2l)$-invariant subspaces within this representation, the ground state of the $\mathrm{SO}(n=2l)$ chain exhibits a two-fold degeneracy.
This MPS formalism naturally explains the fundamental distinction between even and odd $n$ cases -- while odd $n$ chains exhibit a unique ground state, even $n$ chains display a characteristic two-fold degeneracy~\cite{tu2008class}.

In this section, we analyze these exactly solvable points through the lens of parton construction. We establish that the exactly solvable point corresponds to the commuting point of Kitaev chains with parameters $\{\tilde{\chi} = \tilde{\Delta} = 1,~\lambda = 0\}$. Remarkably, the ground state at this point can be exactly represented as a Gutzwiller-projected mean-field state. Using the parton language, we demonstrate that the even-odd distinction arises from $\mathrm{Z}_2\cong\mathrm{O}(n)/\mathrm{SO}(n)$ symmetry breaking rather than translational symmetry breaking in the SO$(n)$ spin chain. Under open boundary conditions, the corresponding zero-energy boundary modes are completely constructed from the Majorana zero modes of the constituent Kitaev chains.

\subsection{Corresponding Kitaev chain at the commuting point}

Consider the commuting point of a single Kitaev chain with parameters $\{\tilde{\chi}=\tilde{\Delta}=1,~\lambda=0\}$, where the fermionic Hamiltonian can be expressed as a summation of local quadratic terms, each of which commutes with the others. The Hamiltonian for a flavor-$\alpha$ chain (temporarily omitting the flavor index) reads:
\begin{equation}\label{eq:HK_a}
H_{\mathrm{K}}= \sum_{i=1}^L\left(- a_i^{\dagger} a_{i+1}+ a_i a_{i+1}+\text{h.c.}\right),
\end{equation}
where $L$ represents the chain length. For an SO$(n)$ spin chain with periodic boundary condition, the corresponding boundary conditions for the partons can be either periodic (PBC, $a_1=a_{L+1}$) or anti-periodic (APBC, $a_1=-a_{L+1}$).

The Hamiltonian $H_\text{K}$ can be diagonalized by introducing Majorana fermions $c_i$ and $d_i$ [see Fig.~\ref{fig:fig1}(b)] through the transformation:
\begin{align*}
a_i^{\dagger} = \frac{1}{2}\left(c_i-\ii d_i\right)\quad\mbox{and}\quad
a_i = \frac{1}{2}\left(c_i+\ii d_i\right).
\end{align*}
In this Majorana representation, the Hamiltonian is re-expressed as:
\begin{equation}\label{eq:HK_Majorana}
H_{\mathrm{K}} = (-1)^{\zeta} \left(\ii d_{L} c_{1}\right) + \sum_{i=1}^{L-1} \left(\ii d_{i} c_{i+1}\right),
\end{equation}
where $\zeta=0$ for PBC and $\zeta=1$ for APBC.

Notably, all bond terms $\ii d_{i}c_{i+1}$ commute with each other. As pictured in Fig.~\ref{fig:fig1}~(b), we can further define bond fermions:
\begin{align*}
f_j^{\dagger} = \frac{1}{2}\left(d_{i-1}+\ii c_i\right)\quad\mbox{and}\quad
f_j = \frac{1}{2}\left(d_{i-1}-\ii c_i\right),
\end{align*}
for $j=2$ to $L+1$, which diagonalize the Hamiltonian up to a constant:
\begin{equation}\label{eq:HK_f}
H_{\mathrm{K}}/2 = -(-1)^{\zeta}f_{L+1}^\dagger f_{L+1} -\sum_{j=2}^{L} f_j^\dagger f_j + \text{const}.
\end{equation}

The ground state for flavor $\alpha$ with PBC ($\zeta=0$) is:
\begin{equation}\label{eq:Kitaev_GS_prod_f}
\ket{\Psi_\mathrm{K}^{\alpha},\zeta=0} = \prod_{j=2}^{L+1} f_{j\alpha}^\dagger \ket{\mathrm{vac}}_f,
\end{equation}
while the APBC ($\zeta=1$) ground state is:
\begin{equation}\label{eq:Kitaev_GS_prod_fA}
\ket{\Psi_\mathrm{K}^{\alpha},\zeta=1} = f_{L+1,\alpha}\ket{\Psi_\mathrm{K}^{\alpha},\zeta=0}.
\end{equation}

Each Kitaev chain preserves a Z$_2$ fermion parity symmetry generated by:
\begin{equation}\label{eq:HK_majorana}
\mathbf{U}_\alpha = \prod_{i=1}^L (-1)^{a_{i\alpha}^\dagger a_{i\alpha}} = \prod_{i=1}^L \left(-\ii c_{i\alpha}d_{i\alpha}\right).
\end{equation}
The ground state parity is determined by the boundary condition:
\begin{equation}\label{eq:parity_of_kitaev_chain}
\mathbf{U}_\alpha \ket{\Psi_\mathrm{K}^{\alpha},\zeta} = (-1)^{\zeta+1} \ket{\Psi_\mathrm{K}^{\alpha},\zeta},
\end{equation}
showing that PBC yields odd parity ($\zeta=0$) while APBC yields even parity ($\zeta=1$).

\begin{figure*}[tb]
\centering
\includegraphics[width=1.0\linewidth]{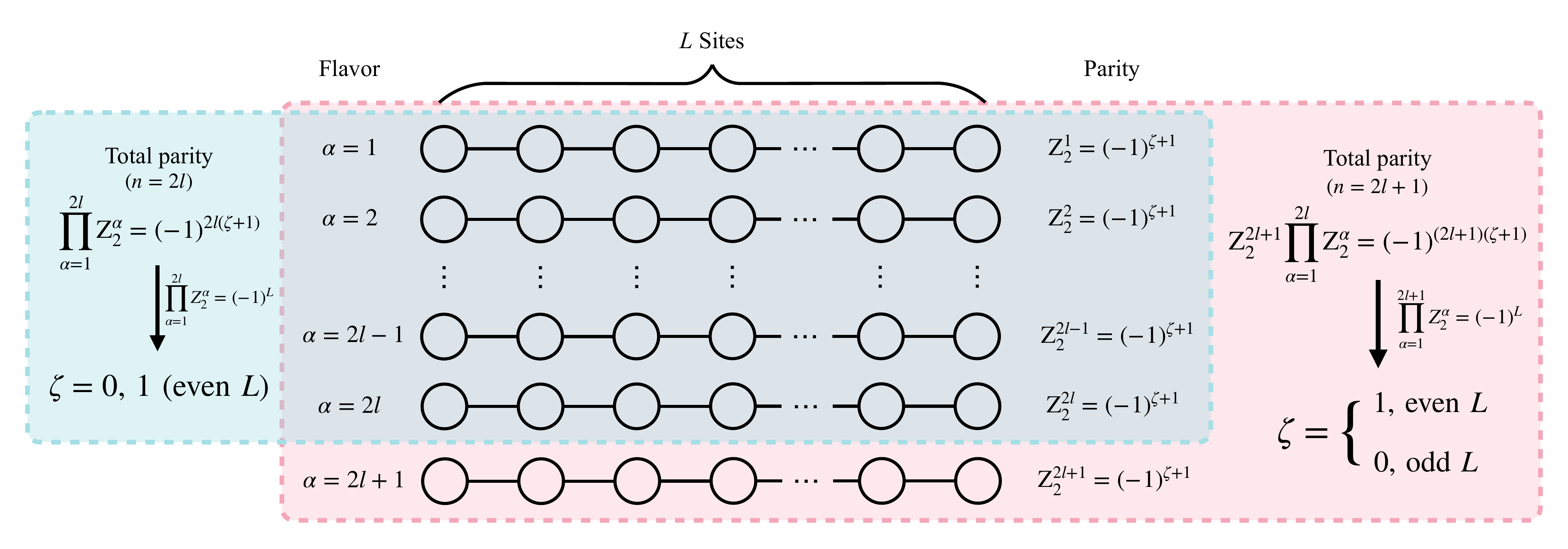}
\caption{The total fermion parity of $\ket{\Psi_{\mathrm{MF}}^\zeta}$ and the ground state degeneracy at the MPS exactly solvable point when $n=2l$ and $n=2l+1$. For a given periodic boundary condition (either APBC or PBC) of the mean-field Hamiltonian, all component Kitaev chains share the same boundary condition label $\zeta$. }
\label{fig:parity}
\end{figure*}

\subsection{Ground state degeneracy: $\mathrm{SO}(2l+1)$ vs. $\mathrm{SO}(2l)$}

The connection between the $\mathrm{SO}(n)$ spin chains and Kitaev chains becomes particularly clear when examining the MPS exactly solvable points. As demonstrated in Refs.~\cite{liu2012gutzwiller}, the Gutzwiller-projected mean-field ground state wave function at $\tilde{\chi} = \tilde{\Delta} = 1$, $\lambda = 0$ exactly corresponds to the MPS ground state wave function of the $\mathrm{SO}(n)$ spin chain:
\begin{equation}\label{eq:PG_mf_to_aklt}
    \mathbf{P}_\mathrm{G} \ket{\Psi_{\mathrm{MF}}(\tilde{\chi}=\tilde{\Delta}=1,~\lambda=0)} = \ket{\Psi_{\text{MPS}}}.
\end{equation}
This remarkable correspondence holds for all $n \geq 3$. The degeneracy of the MPS ground state can be directly understood at the mean-field level. For each flavor $\alpha$, there exist two topologically distinct mean-field ground states $\ket{\Psi_{\mathrm{K}}^\alpha,\zeta=0,1}$ corresponding to PBC and APBC boundary conditions. Both can be Gutzwiller-projected to form trial ground states:
\begin{equation}\label{eq:psi_MF_zeta_01}
\ket{\Psi_{\mathrm{MPS}}^{\zeta}} = \mathbf{P}_{\mathrm{G}}\ket{\Psi_{\mathrm{MF}}^{\zeta}} = \mathbf{P}_{\mathrm{G}}\prod_{\alpha=1}^n \ket{\Psi_{\mathrm{K}}^\alpha,\zeta},
\end{equation}
where $\zeta=0$ (PBC) and $\zeta=1$ (APBC) lead to fundamentally different physical outcomes depending on whether $n$ is odd or even. 
    
\begin{enumerate}
\item{}\textbf{Unique ground state for $n=2l+1$:} For odd $n = 2l+1$, we find that $\ket{\Psi_{\mathrm{MF}}^{\zeta=0}}$ has total parity $(-1)^{2l+1}=-1$ while $\ket{\Psi_{\mathrm{MF}}^{\zeta=1}}$ has total parity $1$. The Gutzwiller projection imposes a parity constraint of $(-1)^L$, leading to qualitatively different behavior for even and odd chain lengths $L$:
\begin{itemize}
\item{} For even $L$: The state $\ket{\Psi_{\mathrm{MF}}^{\zeta=0}}$ with odd parity is eliminated by the projection;
\item{} For odd $L$: The state $\ket{\Psi_{\mathrm{MF}}^{\zeta=1}}$ with even parity is projected out.
\end{itemize}
Consequently, only one mean-field ground state survives the Gutzwiller projection in each case, resulting in a unique ground state for the $\mathrm{SO}(2l+1)$ spin chain at the MPS exactly solvable point.

\item{}\textbf{Two-fold degeneracy and Z$_2$ symmetry for $n=2l$:} For even $n = 2l$, both $\ket{\Psi_{\mathrm{MF}}^{\zeta=0}}$ and $\ket{\Psi_{\mathrm{MF}}^{\zeta=1}}$ possess the same total parity. When $L$ is even, the Gutzwiller projection (requiring total parity $1$) permits both mean-field ground states, leading to a two-fold degeneracy. However, when $L$ is odd, both states are eliminated by the odd total parity constraint.

Contrary to previous suggestions~\cite{tu2008class} that this degeneracy stems from spontaneous translational symmetry breaking, our parton construction reveals that these two ground states actually preserve translational symmetry~\footnote{ Indeed, for a system with physical degrees of freedom under PBC, where the fermionic partons are subject to either PBC or APBC, translational symmetry accounts for the two-fold degeneracy in the $n=2l$ case. However, these degenerate ground states, while linked by translational symmetry, do not themselves respect this symmetry and must be expressed as a linear combination of two MPSs, one with momentum $k=0$ and the other with $k=\pi$, posing challenges for numerical computation using DMRG.}. Instead, the degeneracy originates from Z$_2$ fermion number parity symmetry breaking of individual flavors. The projected ground states satisfy:
\begin{equation*}
\mathbf{U}_{\alpha} \ket{\Psi_{\mathrm{MPS}}^\zeta} = \mathbf{P}_{\mathrm{G}} \mathbf{U}_{\alpha} \ket{\Psi_{\mathrm{MF}}^{\zeta}}=(-1)^{\zeta+1} \ket{\Psi_{\mathrm{MPS}}^\zeta},
\end{equation*}
where the $\alpha$-flavor fermion number parity operator $\mathbf{U}_{\alpha}$ is defined in Eq.~\eqref{eq:HK_majorana}, and the second equality in the equation above is obtained through the application of Eq.~\eqref{eq:parity_of_kitaev_chain}. 
Note that implementing $\mathbf{U}_{\alpha}$ to $a^\dagger_{i\alpha}\rightarrow-a^\dagger_{i\alpha}\,(i=1,\cdots,L)$, while leaves other flavors unchanged. Thereby, in the physical SO($n$) spin basis, this global Z$_2$ transformation $\mathbf{U}_{\alpha}$ corresponds to the $\mathrm{Z}_2\cong\mathrm{O}(n)/\mathrm{SO}(n)$ symmetry operation. 
This demonstrates that the two ground states are distinguished by their eigenvalues of the global $\mathrm{Z}_2\cong\mathrm{O}(n)/\mathrm{SO}(n)$ symmetry, rather than by local order parameters. 

\end{enumerate}

\subsection{Degeneracy under open boundary conditions}

\begin{figure*}[tb]
\centering
\includegraphics[width=1\linewidth]{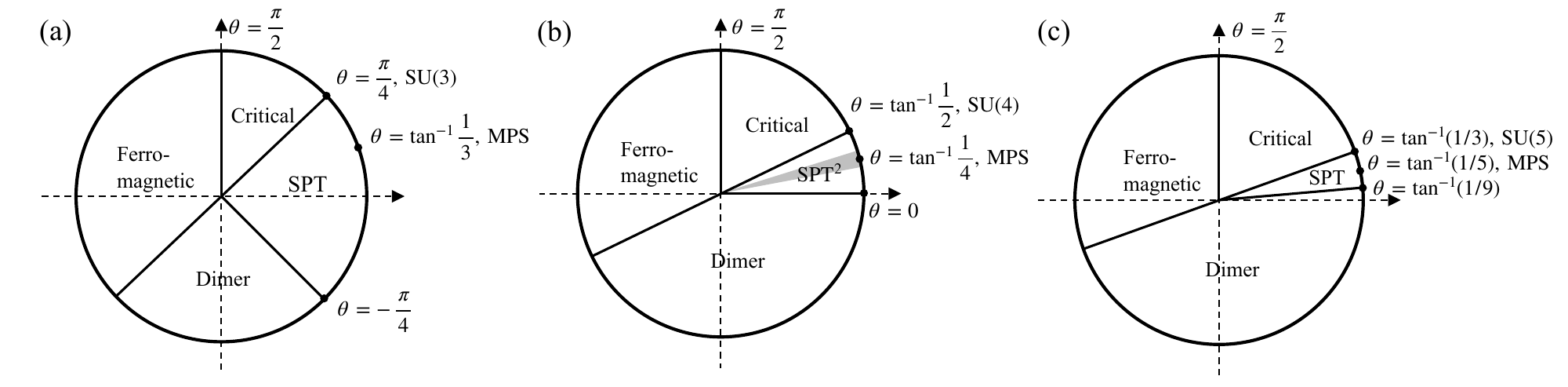}
\caption{The predicted ground state phase diagrams of (a) $\mathrm{SO}(3)$, (b) $\mathrm{SO}(4)$ and (c) $\mathrm{SO}(5)$ BBQ models. The shaded area in (b) corresponds to $0.24\leq \tan\theta\leq 0.26$. }
\label{fig:phasediagrams}
\end{figure*}

Under open boundary condition (OBC), the absence of boundary terms $\sim{}f^\dagger_{L+1,\alpha}f_{L+1,\alpha}$ in $H^\alpha_{\mathrm{K}}$ leads to the emergence of two unpaired Majorana modes $d_{L\alpha}$ and $c_{1\alpha}$ per flavor $\alpha$. This enlarges ground-state degeneracy by introducing boundary Majorana zero modes in addition to the bulk modes. We define the bulk ground state with all finite-energy bulk modes occupied, which takes the form: 
\begin{equation}
\ket{\Psi_{\mathrm{MF},\mathrm{bulk}}} = \prod_{\alpha=1}^{n}\ket{\Psi_{\mathrm{K},\mathrm{bulk}}^\alpha} = \prod_{\alpha=1}^{n}\prod_{j=2}^{L}f_{j\alpha}^\dagger \ket{\mathrm{vac}}_f.
\end{equation}
Accordingly, the parity operator $\mathbf{U}_\alpha$ acts on the bulk state as
\begin{equation}
\mathbf{U}_\alpha \ket{\Psi_{\mathrm{K},\mathrm{bulk}}^\alpha}=(-1)^{L-1}\ket{\Psi_{\mathrm{K},\mathrm{bulk}}^\alpha}.
\end{equation}
Evaluating the total fermion parity of $\ket{\Psi_{\mathrm{MF},\mathrm{bulk}}}$ gives
\[
\mathbf{U}^{\mathrm{tot}}_a \ket{\Psi_{\mathrm{MF},\mathrm{bulk}}} = (-1)^{n(L-1)} \ket{\Psi_{\mathrm{MF},\mathrm{bulk}}}.  
\]  

At the mean-field level, there are $2^n$ distinct  occupation configurations for the boundary zero modes, which results in  the $2^n$-fold degeneracy of the mean-field ground state. However, the Gutzwiller projection, which enforces the $(-1)^L$ parity constraint, reduces this $2^n$ degeneracy to $2^{n-1}$. 
Therefore, the ground-state degeneracy of SO$(n)$ chains under OBC is generally $2^{n-1}$, regardless of whether $n$ is even or odd.
This exactly matches the known degeneracy of the $\mathrm{SO}(n)$ spin chain under OBC. 

Finally, we emphasize that the $\mathrm{Z}_2$ symmetry defined by the $\mathbf{U}_{\alpha}$ operator provides a unified framework that accounts for both bulk and boundary effects under both PBC and OBC, as well as the even-odd effect of $n$, in explaining the ground degeneracy, rather than attributing it to broken translational symmetry.

\section{Beyond the MPS exactly solvable point} \label{sec:results}

While the SO($n$) BBQ chains admit exact MPS solutions at $\tan\theta=1/n$~\cite{tu2008class}, the general case requires comprehensive numerical investigation. In this section, we systematically explore the phase diagram using advanced computational techniques, namely, the  ``Gutzwiller-guided DMRG" method~\cite{jin2021density}. 
Our approach integrates analytical insights with state-of-the-art numerical methods to provide a thorough understanding of these quantum phases.

As demonstrated in Section~\ref{sec:model}, the analytical parton construction formalism reveals that both the topological nature and the even–odd effect persist throughout the entire SPT phase ($n=2l+1$) and/or SPT$^2$ ($n=2l$) phase, not just at the MPS exactly solvable point. In other words, the Gutzwiller-projected parton state given by Eq.~\eqref{eq:mf_to_trail} remains a promising trial wave function for several important phases of the SO($n$) BBQ chains. 
Thereby, we employ numerical approaches to the SO$(n)$ model, building on the following analytical parton construction formalism:
\begin{enumerate}
\item{} Construct mean-field ground states using the MPO-MPS conversion technique~\cite{jin2020efficient, wu2020tensor};
\item{} Implement Gutzwiller projection to enforce the physical Hilbert space constraints;
\item{} Optimize MPS wave funtcions via a two-site DMRG algorithm that strictly imposes the $\prod_{k=1}^l \otimes{}\mathrm{U}(1)$ symmetry, which arises from the $l$-independent generators in the corresponding Cartan sub-algebra, in both $\mathrm{SO}(n=2l)$ and $\mathrm{SO}(n=2l+1)$ cases; 
\item{} Perform systematic finite-size scaling analysis.
\end{enumerate}

Fig.~\ref{fig:phasediagrams} presents the ground-state phase diagrams for the SO$(n)$ model, as defined in Eq.~\eqref{eq:HBBQ} and parameterized in Eq.~\eqref{eq:JKtheta}, for $n = 3$, 4, and 5. 
For both even and odd $n$, the symmetry-protected topological (SPT or SPT$^2$) phase is bounded by two critical points.
The first is the ULS point at $\tan\theta=1/(n-2)$, where, as indicated in Eq.~\eqref{HBBQ_in_chidel}, an enlarged SU$(n)$ symmetry emerges~\cite{uimin1970one, lai1974lattice, sutherland1975model}. The second is the Takhtajan-Babujian point~\cite{takhtajan1982picture, babujian1982exact} for $n=3$ and Reshetikhin point~\cite{reshetikhin1983method, reshetikhin1985integrable} for $n>3$, located at $\tan\theta=(n-4)/(n-2)^2$. In both cases, the SO$(n)$ model is exactly solvable using the Bethe Ansatz. 

The remainder of this section focuses on characterizing the SPT$^2$ phase in SO($n=2l$) chains and the two critical points for arbitrary $n$.

\begin{figure*}[tb]
	\centering
	\includegraphics[width=0.49\linewidth]{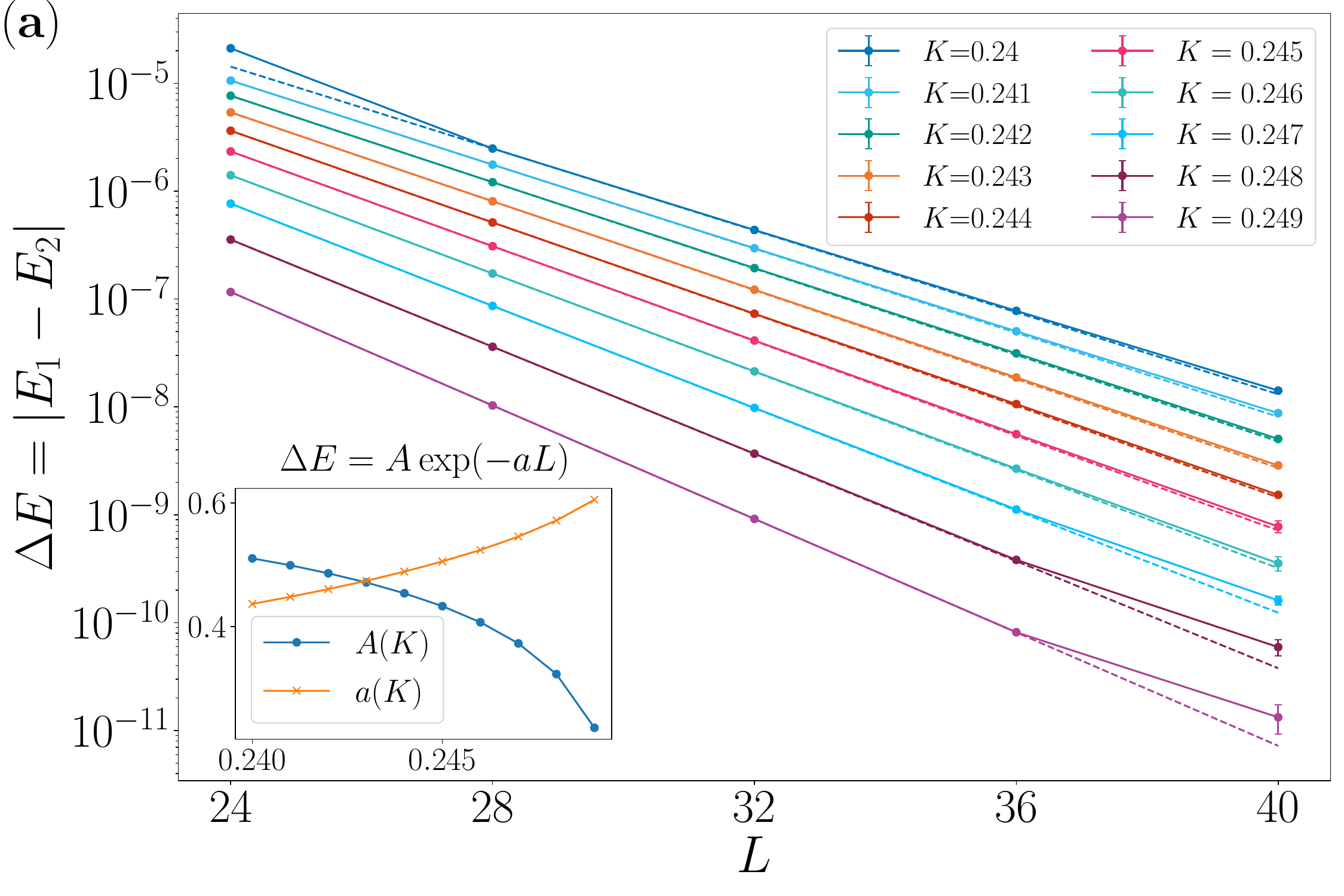}
	\includegraphics[width=0.49\linewidth]{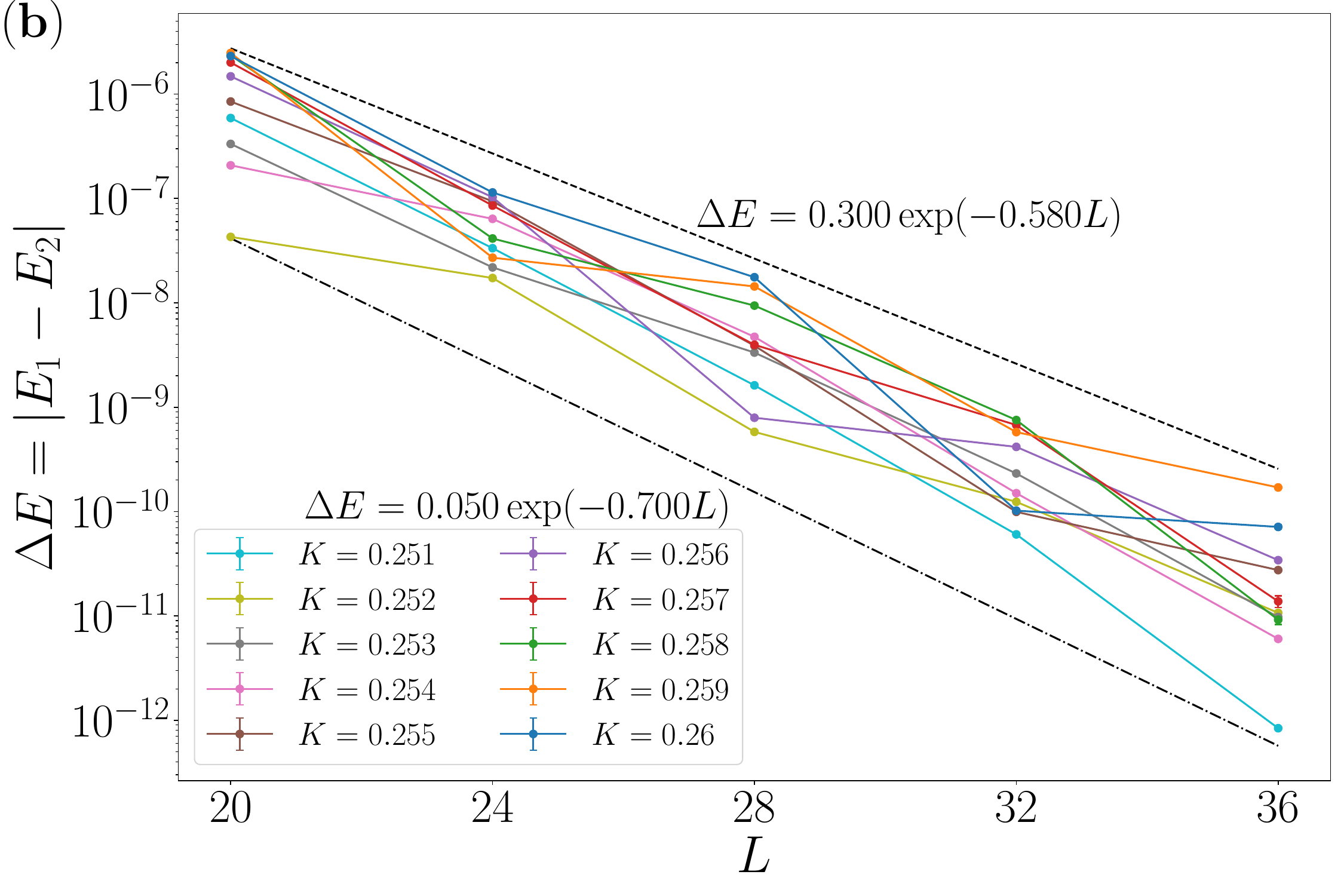}
	\caption{SO($n=4$) model: The energy gap between the two topologically degenerate states, $\Delta E$, as a function of chain length $L$ is shown for (a) $K<0.25$ with $D=2000$ and (b) $K>0.25$ with $D=4000$. The chain lengths are chosen as multiples of $n$ when $n$ is even and the PBC is chosen for the spin chain. 
    Error bars are estimated from the energy difference in the final several DMRG steps, and truncation errors are maintained below $10^{-9}$ throughout the DMRG process. The dashed lines in (a) represent the best-fit exponential decay of the energy gap, $\Delta E=A\exp(-aL)$.
	The inset in (a) displays the fitting parameters $A$ and $a$ as functions of the model parameter $K$.}
	\label{fig:energygaps}
\end{figure*}

\subsection{Topological Degeneracy in the SPT$^2$ Phase}

In this subsection, we focus on the $n=2l$ case to investigate two distinct topological sectors in the SPT$^2$ phase. As discussed in Section ~\ref{sec:MPSpoint}, the ground states in these sectors, characterized by different $\mathrm{Z}_2\cong\mathrm{O}(n)/\mathrm{SO}(n)$ number, are exactly degenerate and locally indistinguishable at the MPS point. We now aim to determine whether this exact degeneracy arises intrinsically from translational symmetry breaking or if it is merely accidental, evolving into topological degeneracy away from the MPS point. To this end, we study the energies of the two sectors in the vicinity of the MPS point using the ``Gutzwiller-guided DMRG'' method. For simplicity, and without affecting the main results, we fix $J=1$ and the system length $L$ to be even. 

The $\mathrm{SO}(4)$ BBQ chain model provides the simplest example for even $n$. We vary $K$ within the shaded region near the exactly solvable MPS point [specifically, $0.24\leq K \leq 0.26$; see Fig.~\ref{fig:phasediagrams}(b)] and obtain two optimized ground states corresponding to the distinct parity sectors. Technical details of this numerical method are provided in Appendix~\ref{apdx:GG-DMRG}.

Fig.~\ref{fig:energygaps} illustrates the energy gap between the two parity sectors. When $K<0.25$, the energy gap exhibits a clear exponential decay with increasing $L$~\cite{oshikawa2019} and can be fitted by an exponential function of the form
\[
\Delta E=A(K)\exp\big[-a(K)L\big],
\]
as shown in the inset of Fig.~\ref{fig:energygaps}(a). 
For $K>0.25$, although the numerical results are less definitive, the exponential decay remains evident when compared with two reference exponential functions; see Fig.~\ref{fig:energygaps}(b).
	
From these numerical results, we conclude that in the SPT$^2$ phase for $n=4$, the Gutzwiller-guided DMRG method identifies two distinct topological sectors. The energy gap between the ground states in these sectors decays exponentially with the system size $L$, implying that the states become topologically degenerate in the thermodynamic limit rather than exactly degenerate. This finding indicates that the observed degeneracy is not a consequence of translational symmetry breaking.

This numerical approach can be straightforwardly generalized to any even $n$ ($n\geq 4$). For instance, similar results obtained for the $\mathrm{SO}(6)$ BBQ chain are presented in Appendix~\ref{apdx:so6_energygap}.

\subsection{Precise Characterization of Critical Behavior}

\begin{figure*}[tb]
\centering
\includegraphics[width=1.0\linewidth]{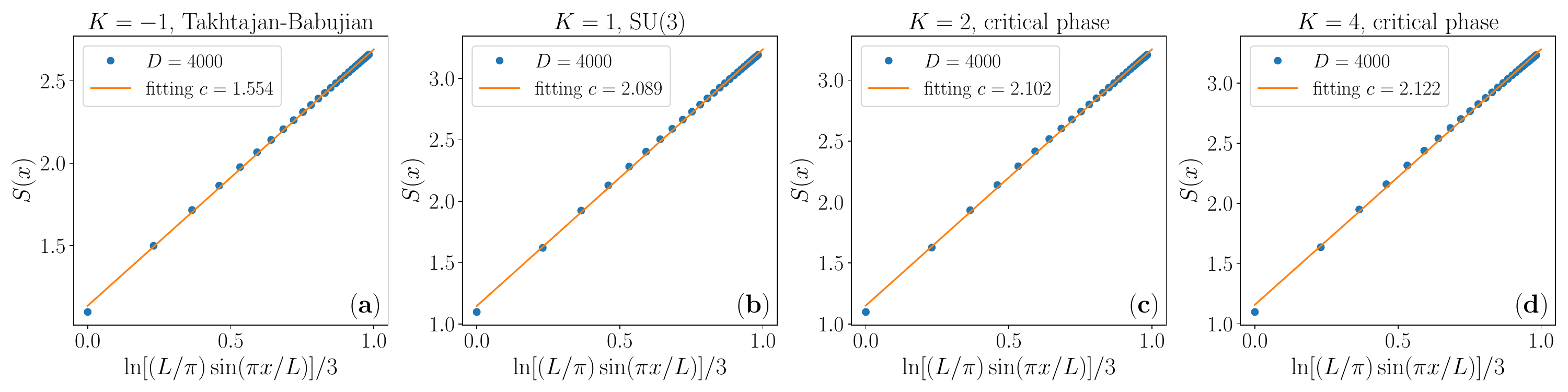}\par
\includegraphics[width=1.0\linewidth]{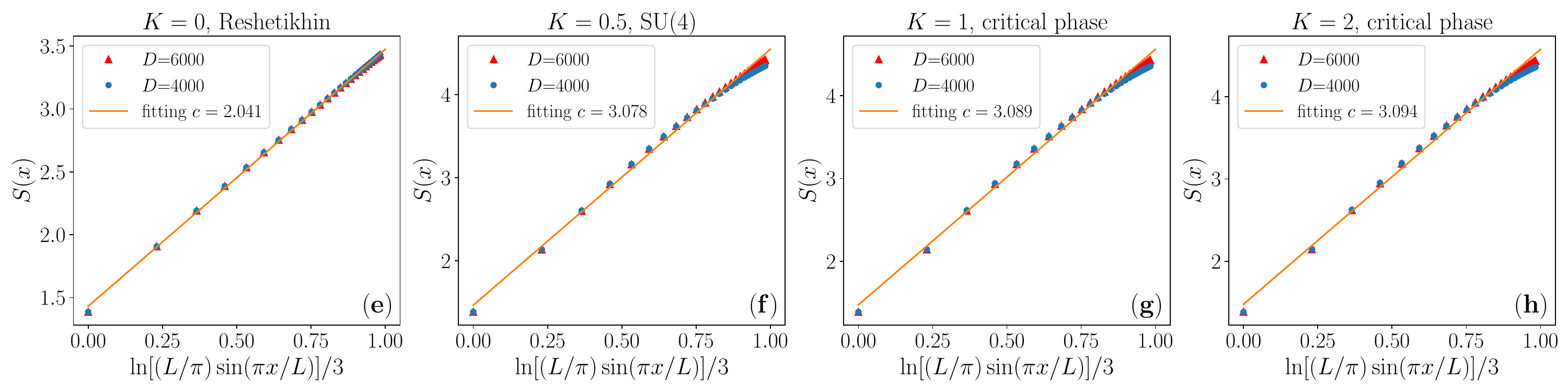}
\caption{Central charge fitting results for (a) -- (d): the $\mathrm{SO}(3)$ and (e) -- (h): the $\mathrm{SO}(4)$ BBQ spin chains, each with a chain length $L=60$ under PBC. Here $J=1$ is fixed while $K$ is varied in Eq.~\eqref{eq:HBBQ}. The DMRG calculations use a maximum bond dimension of $D=4000$ for (a) -- (d) and $D=4000$ as well as $D=6000$ for (e) -- (h). The orange solid lines indicate the fits based on the Calabrese-Cardy formula given in Eq.~\eqref{eq:CC}. In (f) -- (h), only the first 40 data points of the $D=6000$ results are used for the fitting to minimize the influence of the tail.}
\label{fig:ccfitso34}
\end{figure*}

In addition to the gapped SPT/SPT$^2$ phases, several critical points can also be effectively described by Gutzwiller-projected parton states derived from the mean-field Hamiltonian in Eq.~\eqref{eq:HMF}. For example, at the ULS point, the mean-field ground state of $H_{\mathrm{MF}}$ is a gapless Fermi sea with parameters $\{\tilde{\chi}=1,\,\tilde{\Delta}=0,\,\lambda=1\}$, which exactly preserves the enhanced $\mathrm{SU}(n)$ symmetry. At the Takhtajan-Babujian or Reshetikhin point, the mean-field parameters are chosen as $\{\tilde{\chi}=\tilde{\Delta}=1,\,\lambda=2\}$, which indeed, corresponds to the topological phase transition point of the Kitaev chain.
These different critical phases are characterized by their nonzero central charge of the corresponding conformal field theory. 
In practice, the central charge $c$ is extracted by fitting the entanglement entropy with the Calabrese-Cardy formula under PBC~\cite{calabrese2009entanglement}:
\begin{equation}\label{eq:CC}
S(x) = \frac{c}{3} \ln \left[\frac{L}{\pi} \sin \left(\frac{\pi x}{L}\right)\right] + \text{const.},
\end{equation}
where $S(x)$ is the von Neumann entanglement  entropy of a subsystem with size $x$, and $L$ is the total system size.
These parton states serve as the initial states for the “Gutzwiller-guided DMRG” method, which allows us to efficiently extract the central charges of these critical states.

Using the Gutzwiller-guided DMRG, we obtain the MPS representations of the ground states and determine their central charges for several important critical states in the $n=3$ and $n=4$ BBQ chains: (i)  the Takhtajan-Babujian (Reshetikhin) point for $n=3$ ($n=4$), (ii) the SU$(n)$ symmetric ULS point, and (iii) the critical phase in the parameter range of $\tan^{-1}\left[1/(n-2)\right]<\theta<\pi/2$ (see Fig.~\ref{fig:phasediagrams}).
The central charge fitting results, shown in Fig.~\ref{fig:ccfitso34}, reveal the following:
\begin{itemize}
\item{} Takhtajan-Babujian/Reshetikhin point: We find $c = 1.554$ for $n=3$ and $c = 2.041$ for $n=4$, which align with the predictions of the $\mathrm{SO}(n)_1$ Wess-Zumino-Witten (WZW) conformal field theory~\cite{Tsvelik1990,tu2011effective}, where $c=n/2$.
\item{} ULS point: We obtain $c = 2.089$ for $n=3$ and $c = 3.078$ for $n=4$, in agreement with the $\mathrm{SU}(n)_1$ WZW conformal field theory~\cite{Affleck1988}, which predicts $c=n-1$.
\item{} Critical phase: In the critical phase, we further observe:
    \begin{itemize}
        \item $n=3$: $c=2.102$ and $c=2.122$ for $\theta=\tan^{-1}2$ and $\theta=\tan^{-1}4$. 
        \item $n=4$: $c=3.089$ and $c=3.094$ for $\theta=\pi/4$ and $\theta=\tan^{-1}2$. 
    \end{itemize}
\end{itemize}

Notably, as seen in Fig.~\ref{fig:ccfitso34}(f)-(h), tails appear in the entanglement entropy data for $K=0.5$, $K=1$, and $K=2$ as a result of the limited bond dimension. A comparison between the $D=4000$ and $D=6000$ data indicates that increasing the bond dimension gradually reduces this tail. Moreover, the convergence of the half-chain entanglement entropy at the ULS point is significantly slower than that at the Reshetikhin point, which facilitates the emergence of the tail. Consequently, a larger bond dimension is required to accurately capture the half-chain entanglement structure. As an example of the generalization to arbitrary $n$ ($n\geq 3$), similar results for the $\mathrm{SO}(5)$ and $\mathrm{SO}(6)$ chains are presented in Appendix~\ref{apdx:cc_so5_so6}.

\section{Summary and Discussions}\label{sec:sum}

In this work, we have systematically investigated the symmetry-protected topological (SPT or SPT$^2$) phases and critical behavior of SO($n$) BBQ chains using a combination of analytical and numerical approaches. 
Our key findings can be summarized as follows:
\begin{enumerate}
\item{}Topological Degeneracy: Through parton construction and strictly implementing Gutzwiller projection, we have demonstrated that SO($n=2l$) chains exhibit a characteristic two-fold topological degeneracy in the SPT$^2$ phase, while SO($n=2l+1$) chains show a unique ground state. This distinction arises from the different Z$_2$ fermion parity properties of the projected mean-field states, which correspond to the $\mathrm{Z}_2\cong\mathrm{O}(n)/\mathrm{SO}(n)$ symmetry in the physical SO($n$) spin basis.

\item{}Numerical Verification: Using the Gutzwiller-guided DMRG method, we have numerically confirmed that the energy gap between the two different topological sectors in SO($n=4\,\mbox{and}\,6$) chains decays exponentially with system size, establishing their topological nature rather than accidental degeneracy. 

\item{}Critical Behavior: We have characterized several critical points, including the ULS and Reshetikhin points, showing excellent agreement with Wess-Zumino-Witten conformal field theory predictions. The central charges extracted from entanglement entropy scaling match the expected values for both SO($n$)$_1$ and SU($n$)$_1$ theories.
\end{enumerate} 

The implications of our results are several-fold: First, the identification of distinct topological sectors in even-$n$ chains provides new insights into the characterization of SPT phases in high-symmetry systems. 
Several numerical challenges still remain, e.g., the slow convergence of the half-chain entanglement entropy at critical states as the value $n$ increases. These challenges highlight the significance of developing more efficient tensor network algorithms for studying high-symmetry systems.
The success of our Gutzwiller-guided approach suggests that combining analytical insights with numerical methods can significantly enhance our ability to study complex quantum phases.

Our work establishes a firm foundation for further exploration of topological phases in high-symmetry quantum systems, bridging the gap between theoretical predictions and numerical verification. The methods developed here should prove valuable for studying other topological phases in one and two-dimensional systems~\cite{affleck1988valence,jin2025}.

\section*{Acknowledgment}
We would like to express our gratitude to Hong-Hao Tu and Zheng-Xin Liu for their crucial insights and collaborations on numerous pertinent topics. The work is supported in part by National Key Research and Development Program of China (No. 2022YFA1403403), National Natural Science Foundation of China (No.12274441, 12034004). H.-K.J. acknowledges the support from the start-up funding from ShanghaiTech University.

\appendix
\section{Derivation of Eq.~\eqref{HBBQ_in_chidel}}

The key to this derivation lies in understanding the rotation rules of local states, as expressed in Eq.~\eqref{eq:son_rotation}. For notational simplicity, we represent the local state $\ket{m^a}$ as $\ket{\alpha}$ (where $\alpha=1,2,\dots,n$). In the basis $\{\ket{1}, \ket{2}, \dots, \ket{n}\}$, the matrix elements of $\mathbf{L}^{ab}$ are given by:
\begin{equation*}
\left(L^{ab}\right)_{\alpha\beta} = \left\langle \alpha \big| L^{ab} \big| \beta\right\rangle = \ii(\delta_{a\alpha}\delta_{b\beta}-\delta_{a\beta}\delta_{b\alpha}).
\end{equation*}
Consequently, the generator $\mathbf{L}_{i}^{ab}$ can be expressed in terms of parton operators as
\begin{equation}\label{eq:L_in_partons}
\mathbf{L}_i^{ab}=\sum_{\alpha,\beta} a_{i\alpha}^\dagger \left(L^{ab}\right)_{\alpha\beta} a_{i\beta} = \ii\left( a_{ia}^\dagger a_{ib} - a_{ib}^\dagger a_{ia} \right).
\end{equation}

The derivation of Eq.~\eqref{HBBQ_in_chidel} reduces to demonstrating the following claim: For all integers $n\geq 3$, the following relations hold:
\begin{equation*}
\sum_{a<b}^{n} \mathbf{L}_i^{a b} \mathbf{L}_{j}^{a b} = -\left( \hat{\chi}_{ij}^\dagger \hat{\chi}_{ij} + \hat{\Delta}_{ij}^\dagger \hat{\Delta}_{ij} \right) + \sum_{\alpha=1}^n a_{i\alpha}^\dagger a_{i\alpha},
\end{equation*}
and
\begin{equation*}
\left(\sum_{a<b}^{n} \mathbf{L}_i^{a b} \mathbf{L}_{j}^{a b}\right)^2 = (n-2)\hat{\Delta}_{ij}^\dagger \hat{\Delta}_{ij} + \left(\sum_{\alpha=1}^n a_{i\alpha}^\dagger a_{i\alpha}\right)\left(\sum_{\alpha=1}^n a_{j\alpha}^\dagger a_{j\alpha}\right),
\end{equation*}
where, for reminder, $\hat{\chi}_{ij}$ and $\hat{\Delta}_{ij}$ are defined as in Eq.~\eqref{chi_and_del_operators}. Consequently, under the single-occupancy constraint, the BBQ Hamiltonian takes the form:
\begin{equation*}
H = \sum_{\left\langle i,j \right\rangle}\left\{ -J\hat{\chi}_{ij}^\dagger \hat{\chi}_{ij} + \left[\left(n-2\right)K-J\right]\hat{\Delta}_{ij}^\dagger \hat{\Delta}_{ij}+(K+J)\right\}.
\end{equation*}

\textbf{Proof}: We establish this claim through mathematical induction. For the base case $n=3$, as discussed in Section \ref{sec:model} in the main text, the $\mathrm{SO}(3)$ case clearly satisfies:
\begin{equation}\label{so3_1}
	\sum_{a<b}^{n=3} \mathbf{L}_i^{a b} \mathbf{L}_{j}^{a b} = -\left( \hat{\chi}_{ij}^\dagger \hat{\chi}_{ij} + \hat{\Delta}_{ij}^\dagger \hat{\Delta}_{ij} \right) + \sum_{\alpha=1}^n a_{i\alpha}^\dagger a_{i\alpha},
\end{equation}
\begin{equation}\label{so3_2}
	\left(\sum_{a<b}^{n=3} \mathbf{L}_i^{a b} \mathbf{L}_{j}^{a b}\right)^2 = \hat{\Delta}_{ij}^\dagger \hat{\Delta}_{ij} + \left(\sum_{\alpha=1}^n a_{i\alpha}^\dagger a_{i\alpha}\right)\left(\sum_{\alpha=1}^n a_{j\alpha}^\dagger a_{j\alpha}\right),
\end{equation}
and
\begin{equation}\label{so3_3}
	H = \sum_{\left\langle i,j \right\rangle}\left[ -J\hat{\chi}_{ij}^\dagger \hat{\chi}_{ij} + \left(K-J\right)\hat{\Delta}_{ij}^\dagger \hat{\Delta}_{ij} + (K+J)\right].
\end{equation}

\begin{figure*}[t]
\centering
\includegraphics[width=0.49\linewidth]{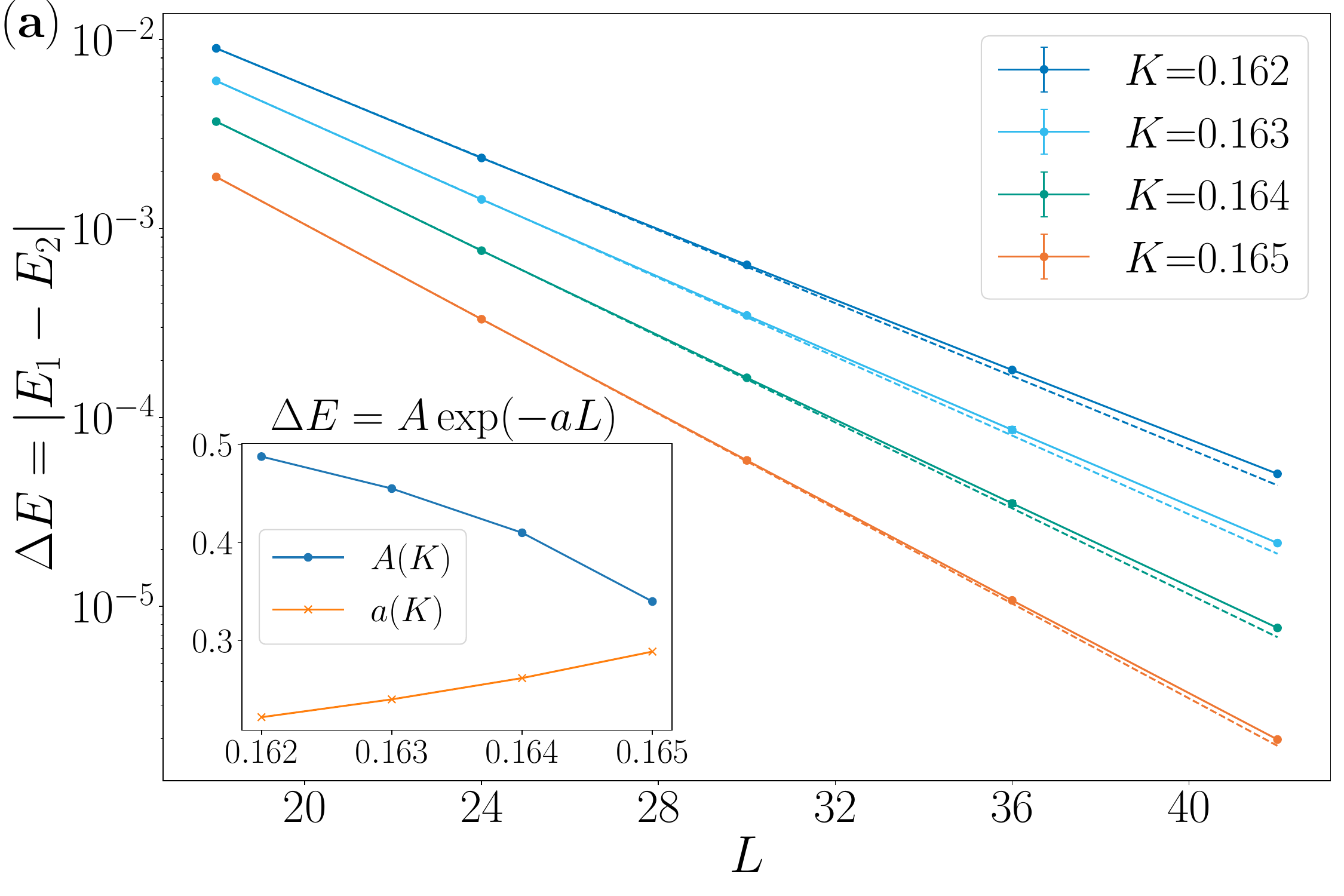}
\includegraphics[width=0.49\linewidth]{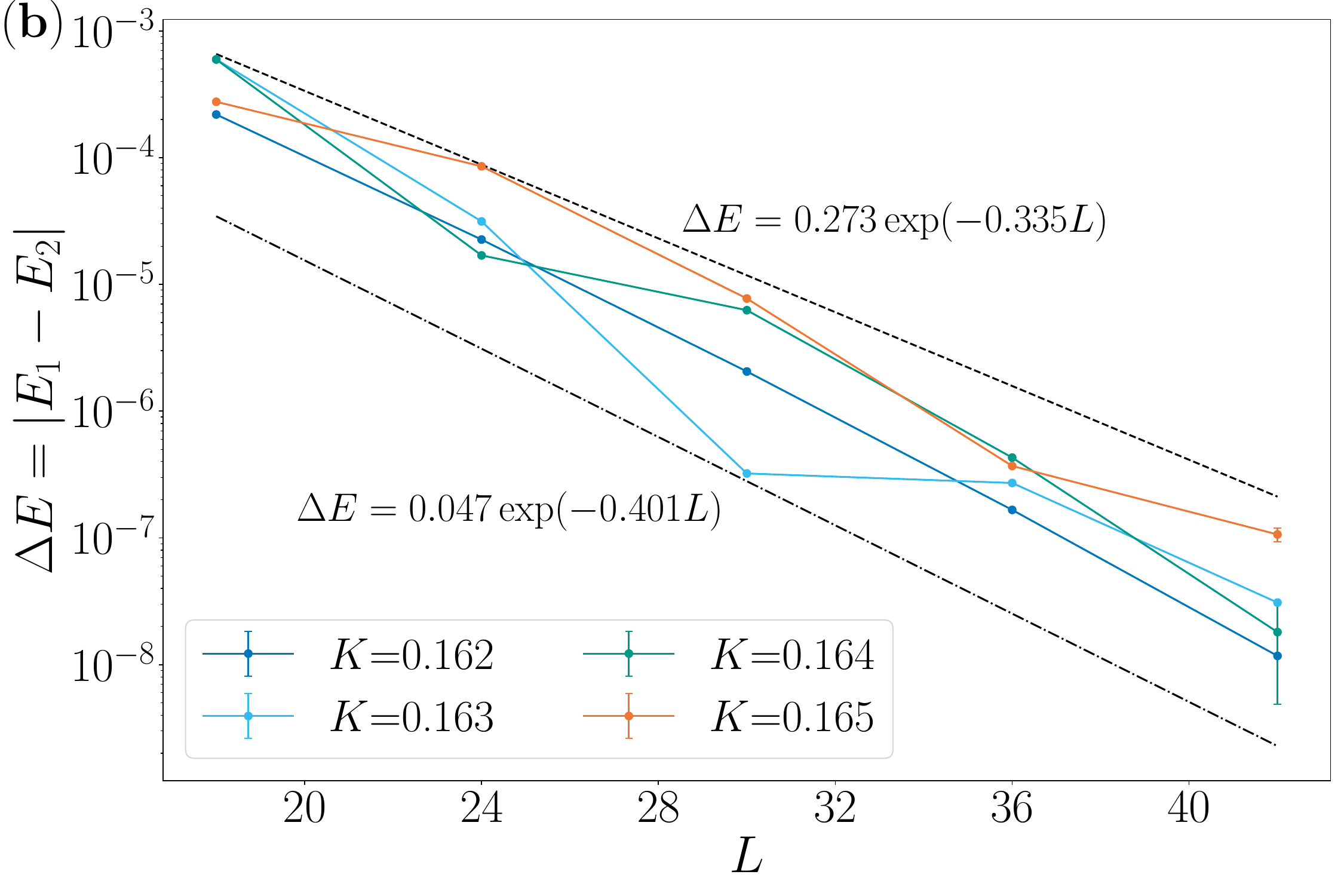}
\caption{$\mathrm{SO}(6)$ model: The energy gap between two topologically degenerate states, $\Delta E$, as a function of chain length $L$ is shown for (a) $K<1/6$ and (b) $K>1/6$, both calculated with bond dimension $D=4000$. Error bars are estimated from the energy difference in the final several DMRG steps, and truncation errors are maintained below $10^{-9}$ throughout the DMRG process. The dashed lines in (a) shows the best-fit exponential decay of the energy gap $\Delta E=A \exp(-aL)$. The inset in (a) displays the fitting parameters $A$ and $a$ as functions of the model parameter $K$. }
\label{fig:energygapsso6}
\end{figure*}

For the inductive step, we have confirmed the validity of Eqs.~\eqref{so3_1}--\eqref{so3_3} for several $n$ ($n>3$) and then proceed to demonstrate their validity for $n+1$. Specifically, if the mathematical induction holds, the following equations must be satisfied:
\begin{equation}\label{claim1}
	\sum_{a<b}^{n+1} \mathbf{L}_i^{a b} \mathbf{L}_{j}^{a b} = -\left( \hat{\chi}_{ij}^\dagger \hat{\chi}_{ij} + \hat{\Delta}_{ij}^\dagger \hat{\Delta}_{ij} \right) + \sum_{\alpha=1}^{n+1} a_{i\alpha}^\dagger a_{i\alpha}
\end{equation}
and
\begin{equation}\label{claim2}
	\left(\sum_{a<b}^{n+1} \mathbf{L}_i^{a b} \mathbf{L}_{j}^{a b}\right)^2 = (n-1)\hat{\Delta}_{ij}^\dagger \hat{\Delta}_{ij} + \left(\sum_{\alpha=1}^{n+1} a_{i\alpha}^\dagger a_{i\alpha}\right)\left(\sum_{\alpha=1}^{n+1} a_{j\alpha}^\dagger a_{j\alpha}\right),
\end{equation}
where $\hat{\chi}_{ij} = \sum_{\alpha=1}^{n+1} a_{i\alpha}^\dagger a_{j\alpha}$ and $\hat{\Delta}_{ij}=\sum_{\alpha=1}^{n+1} a_{i\alpha}a_{j\alpha}$. This would imply that under the single-occupancy constraint, the BBQ Hamiltonian becomes:
\begin{equation}
H = \sum_{\left\langle i,j \right\rangle}\left[ -J\hat{\chi}_{ij}^\dagger \hat{\chi}_{ij} + \left((n-1)K-J\right)\hat{\Delta}_{ij}^\dagger \hat{\Delta}_{ij} + (K+J)\right].
\end{equation}

To distinguish between different cases, we introduce subscripts $n$ and $n+1$ for terms like $\hat{\chi}_{ij}^\dagger\hat{\chi}_{ij}$, denoting them as $\left(\hat{\chi}_{ij}^\dagger\hat{\chi}_{ij}\right)_{n}$ and $\left(\hat{\chi}_{ij}^\dagger\hat{\chi}_{ij}\right)_{n+1}$, respectively. The same convention applies to $\hat{\Delta}_{ij}^\dagger\hat{\Delta}_{ij}$ terms.

The proof of Eqs.~\eqref{claim1} and \eqref{claim2} can be reduced to verifying the validity of the differences between these equations and their $n$-th counterparts. After substituting the operators $\mathbf{L}_{i}^{ab}$, $\hat{\chi}_{ij}$, and $\hat{\Delta}_{ij}$ with their parton operator representations [Eqs.~\eqref{eq:L_in_partons} and \eqref{chi_and_del_operators}], we obtain the difference equations as follows,
\begin{equation}\label{eq:eqclaim1}
\begin{aligned}
    &\sum_{a<b}^{n+1} L_{i}^{ab} L_{j}^{ab} - \sum_{a<b}^{n} L_{i}^{ab} L_{j}^{ab} \\
    =& -\left[\left(\hat{\chi}_{ij}^\dagger\hat{\chi}_{ij}\right)_{n+1} - \left(\hat{\chi}_{ij}^\dagger\hat{\chi}_{ij}\right)_{n} + \left(\hat{\Delta}_{ij}^\dagger\hat{\Delta}_{ij}\right)_{n+1} - \left(\hat{\Delta}_{ij}^\dagger\hat{\Delta}_{ij}\right)_{n}\right] \\
    &+a_{i,n+1}^\dagger a_{i,n+1}
\end{aligned}
\end{equation}
and
\begin{equation}\label{eq:eqclaim2}
\begin{aligned}
&\left(\sum_{a<b}^{n+1} L_i^{a b} L_{j}^{a b}\right)^2 - \left(\sum_{a<b}^{n} L_i^{a b} L_{j}^{a b}\right)^2 \\
=& (n-2)\left[\left(\hat{\Delta}_{ij}^\dagger\hat{\Delta}_{ij}\right)_{n+1} - \left(\hat{\Delta}_{ij}^\dagger\hat{\Delta}_{ij}\right)_{n}\right] + \left(\hat{\Delta}_{ij}^\dagger\hat{\Delta}_{ij}\right)_{n+1} \\
&+ \sum_{\alpha=1}^n a_{i\alpha}^{\dagger} a_{i\alpha} a_{j,n+1}^{\dagger} a_{j,n+1}+a_{i,n+1}^{\dagger} a_{i,n+1} \sum_{\alpha=1}^n a_{j \alpha}^{\dagger} a_{j \alpha} \\
&+a_{i,n+1}^{\dagger} a_{i,n+1} a_{j,n+1}^{\dagger} a_{j,n+1}.
\end{aligned}
\end{equation}
After straightforward algebra, we confirm that Eqs.~\eqref{eq:eqclaim1} and \eqref{eq:eqclaim2} hold, thereby completing the proof.

\section{The Gutzwiller-guided DMRG Method}\label{apdx:GG-DMRG}

The Hamiltonian of a single Kitaev chain can be expressed in Bogoliubov-de Gennes (BdG) form, enabling the application of the MPO-MPS technique for constructing corresponding many-body ground states. The states described by Eqs.~\eqref{eq:Kitaev_GS_prod_f} and \eqref{eq:Kitaev_GS_prod_fA} involve multiple parton creation operators $a_{i\alpha}^\dagger$, which can be efficiently represented as sequences of MPOs~\cite{wu2020tensor,jin2020efficient}. By contracting these MPOs with the $a$-fermion vacuum state, we obtain the many-body ground state in MPS form. This representation serves as an excellent initial state for subsequent DMRG calculations~\cite{jin2021density}.

The Gutzwiller-guided DMRG methodology comprises the following steps:

\begin{enumerate}
\item{} Select appropriate parameters $\{\tilde{\chi}, \tilde{\Delta}, \lambda\}$ for the mean-field Hamiltonian $H_{\mathrm{MF}}$. For instance, at the MPS exactly solvable point, we choose $\tilde{\chi}=1$, $\tilde{\Delta}=1$, and $\lambda=0$.
	
\item{} Generate mean-field ground states for both PBC and APBC conditions using the MPO-MPS method, denoted as $\ket{\Psi_{\mathrm{MF}}^{\zeta=1}}$ and $\ket{\Psi_{\mathrm{MF}}^{\zeta=0}}$, respectively.
	
\item{} Apply the Gutzwiller projection operator to obtain trial states:
\begin{align*}
\ket{\Psi_{\mathrm{trial}}^{1}} &= \mathbf{P}_{\mathrm{G}}\ket{\Psi_{\mathrm{MF}}^{\zeta=1}} \\
\ket{\Psi_{\mathrm{trial}}^{2}} &= \mathbf{P}_{\mathrm{G}}\ket{\Psi_{\mathrm{MF}}^{\zeta=0}}
\end{align*}
These projected states serve as initial conditions for DMRG optimization.
	
\item{} With $J=1$ fixed, perform two-site DMRG calculations for each value of $K$ to optimize ground states of the BBQ Hamiltonian given in Eq.~\ref{eq:HBBQ}. For each trial state, conduct independent DMRG runs, producing two distinct optimized MPSs ($\ket{\psi_1}$ and $\ket{\psi_2}$) with final bond dimension $D$.
	
\item{} Construct and analyze the following matrices:
\begin{align*}
A &= \begin{pmatrix}
\langle\psi_1|H|\psi_1\rangle & \langle\psi_1|H|\psi_2\rangle \\
\langle\psi_2|H|\psi_1\rangle & \langle\psi_2|H|\psi_2\rangle
\end{pmatrix} \\
B &= \begin{pmatrix}
\langle\psi_1|\psi_1\rangle & \langle\psi_1|\psi_2\rangle \\
\langle\psi_2|\psi_1\rangle & \langle\psi_2|\psi_2\rangle
\end{pmatrix},
\end{align*}
where $H$ is defined in Eq.~\eqref{eq:HBBQ}.
Solve the generalized eigenvalue problem $A\mathbf{v}=EB\mathbf{v}$ to obtain eigenvalues $E_1$ and $E_2$, corresponding to ground state energies for distinct Z$_2$ sectors of O$(n)$/SO($n$) symmetry.
\end{enumerate}

\begin{figure*}[tb]
\centering
\includegraphics[width=1.0\linewidth]{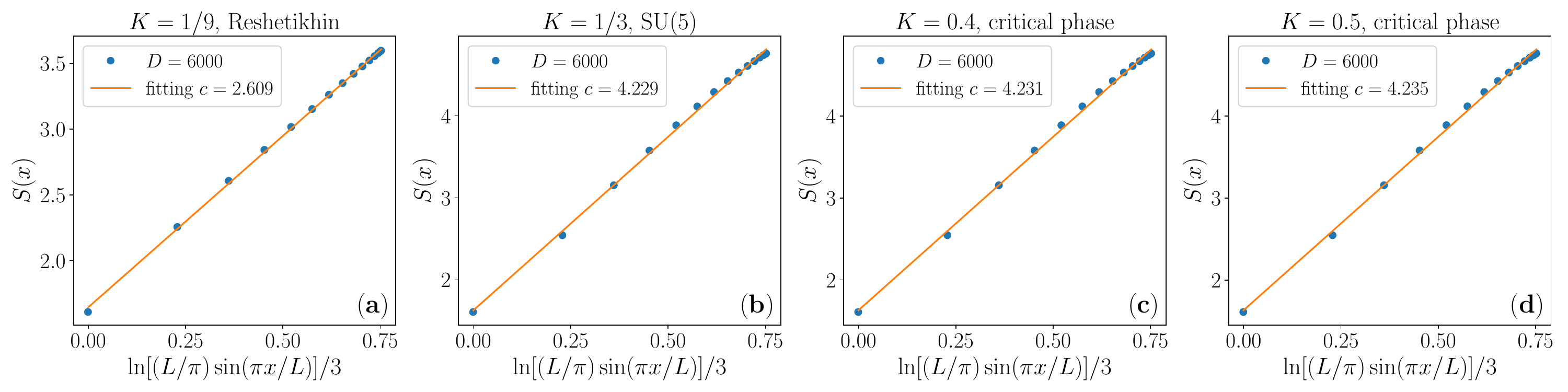}
\includegraphics[width=1.0\linewidth]{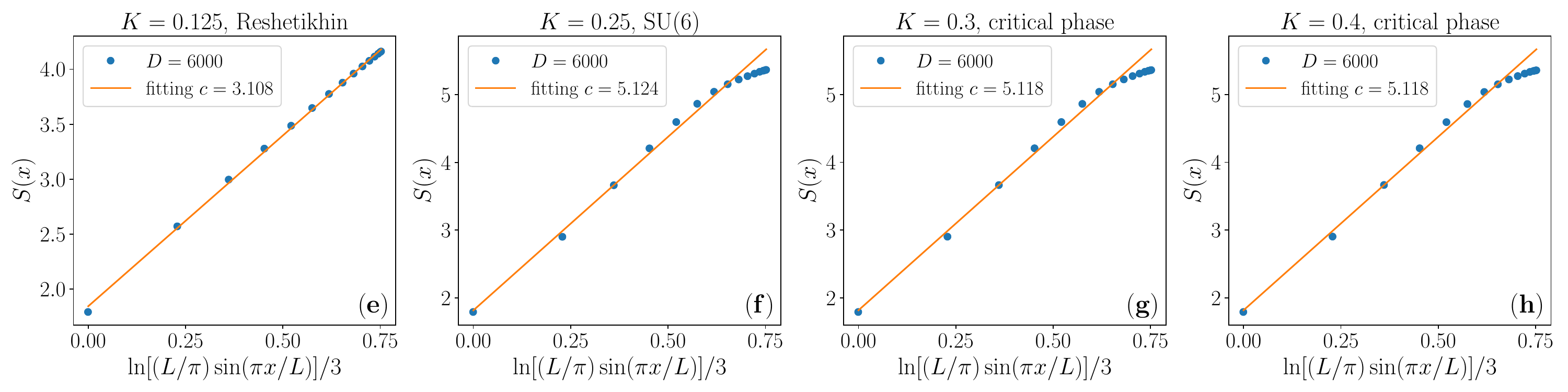}
\caption{Central charge fitting results for (a) -- (d): the $\mathrm{SO}(5)$ and (e) -- (h): the $\mathrm{SO}(6)$ BBQ spin chains, each with chain length $L=30$ under PBC. $J=1$ is fixed while $K$ is varied in Eq.~\eqref{eq:HBBQ}. The DMRG calculations use a maximum bond dimension $D=6000$. The orange solid lines indicate the fits based on the Calabrese-Cardy formula. In (f) -- (h), only the first 20 data points are used for the fitting to minimize influence of the tail.}
\label{fig:ccfitso56}
\end{figure*}

\section{Topological Degeneracy in the $\mathrm{SO}(6)$ Model}\label{apdx:so6_energygap}

The two-fold degenerate ground states are also observed at the exactly solvable MPS point $K=1/6$ of the SO(6) BBQ chain. 
The SPT$^2$ phase in the SO(6) BBQ chain is sandwiched between the Reshetikhin point at $K = 1/8$ and the ULS point at $K=1/4$. It is worth noting that as $n$ increases, the SPT phase region shrinks. Thereby, determining the exponentially decaying energy gap between topologically distinct ground-state sectors becomes more challenging as $n=2l$ being larger.

Our findings for the $n=6$ case are similar to those for $n=4$ in the main text. Fig.~\ref{fig:energygapsso6}(a) demonstrates clear exponential behavior on the side neighboring the dimerized phase ($K<1/6$) in the SPT$^2$ phase. Conversely, Fig.~\ref{fig:energygapsso6}(b) reveals that energy gaps on the side neighboring the  critical phase ($K>1/6$) follow two distinct trajectories in the semi-logarithmic plot, confirming their exponential decay characteristics.

\section{Critical Behaviors in $\mathrm{SO}(5)$ and $\mathrm{SO}(6)$ Models}\label{apdx:cc_so5_so6}

We numerical study the central charges at the Reshetikhin points and ULS points of the SO$(n=5)$ and SO$(n=6)$ points, as shown in Fig.~\ref{fig:ccfitso56}. 
Two additional points within the critical phase are also examined. All of the results are consistent with the predictions given by the conformal field theory~\cite{Tsvelik1990,tu2011effective,Affleck1988}. 
However, the analysis becomes progressively more demanding as $n$ increases, due to fundamental computational constraints. 
From a technical standpoint, the $\mathrm{SO}(n)$ group's $n(n-1)/2$ generators lead to a quadratic growth in local operators, significantly increasing numerical complexity. This scaling requires substantially higher bond dimensions to properly capture half-chain entanglement entropy as $n$ grows. For instance, the DMRG calculations for SO$(6)$ cases are implemented with parameters $L=30$ and $D=6000$. 

We would like to emphasize that, unlike the gapped systems discussed in the main text, MPO-MPS initialization provides no clear advantage over random initialization for these critical systems~\footnote{In these critical systems, (1) the ground state is unique, (2) the entanglement entropy grows only logarithmically, and (3) the variational landscape encountered by DMRG is essentially convex. Because of these features, both a random initial MPS and a Gutzwiller-projected seed reach the same critical fixed point with comparable CPU time; hence the practical benefit of a Gutzwiller warm-start disappears for that very specific situation.}. Consequently, we present results obtained from randomly initialized DMRG rather than the Gutzwiller-guided approach used previously.

\bibliography{SONSPT}

\end{document}